\documentclass{aa}
\usepackage{hyperref}
\usepackage{graphicx}
\usepackage{natbib  }
\usepackage{color   }
\usepackage{array   }
\usepackage{txfonts }
\usepackage{breakurl}
\usepackage{multirow}
\usepackage{multicol}
\usepackage{booktabs}
\usepackage{dcolumn}

 \def\Jyb{\textrm{Jy~beam}$^{-1}$}        % rms noise: Jy/beam
 \def\am{$^\prime$}                       % arcmin
 \def\as{$^{\prime\prime}$}               % arcsec
 \def\Msun{$\mathrm{M}_\odot$}            % solar mass
 \newcommand{\cm}[1]{$\mathrm{cm}^{#1}$}  % absorbing column/particle density
 \newcommand{\dms}[3]{$#1^{#2}\!\!.#3$}   % deg/min/sec and decimal point vertically aligned
 \def\Fermi{{\slshape Fermi}-LAT}         % Fermi-LAT telescope's name
 \def\kes{Kes~41}                         % 'Kes 41'
 \def\Gkes{G337.8$-$0.1}                  % 'G' name of SNR Kes 41

\begin{document} 

\title{A revisited study of the unidentified $\gamma$-ray emission towards the SNR~Kes~41}

\titlerunning{Modelling the $\gamma$-ray emitting region towards SNR~{\kes}}

\author{L. Supan\inst{1,2}
   \and G. Castelletti \inst{1,2}
   \and A. D. Supanitsky \inst{1,2}
   \and M. G. Burton \inst{3,4}
}
\authorrunning{Supan et al.}

\offprints{L. Supan}
\institute {CONICET-Universidad de Buenos Aires, Instituto de Astronomía y Física del Espacio (IAFE), Buenos Aires, Argentina
\and Universidad de Buenos Aires, Facultad de Ciencias Exactas y Naturales. Buenos Aires, Argentina \\
     \email{lsupan@iafe.uba.ar}
\and School of Physics, University of New South Wales, Sydney, NSW 2052, Australia 
\and Armagh Observatory and Planetarium, College Hill, Armagh BT61 9DG, UK }

\date{Received 9 April 2018 / Accepted 13 August 2018}

\abstract
{{\kes} is among the Galactic supernova remnants (SNRs) proposed to be physically linked to $\gamma$-ray emission at GeV energies. Although not conclusively, the nature of the $\gamma$-ray photons has been explained by means of hadronic collisions of particles accelerated at the SNR blast wave with target protons in an adjacent molecular clump. 
We performed an analysis of \emph{Fermi}-Large Area Telescope (LAT) data of about 9 years to assess the origin of the $\gamma$-ray emission. To investigate this matter we also used spectral modelling constraints from the physical properties of the interstellar medium towards the $\gamma$-ray emitting region along with a revised radio continuum spectrum of {\kes} ($\alpha = -0.54 \pm 0.10$, $S\propto\nu^{\alpha}$). 
We demonstrated that the $\gamma$-ray fluxes in the GeV range can be explained through bremsstrahlung emission from electrons interacting with the surrounding medium. We also consider a model in which the emission is produced by pion-decay after hadronic collisions, and confirmed that this mechanism cannot be excluded.}


\maketitle

\section{Introduction}
Molecular clouds (MCs) house stellar objects at different stages of their evolution, from star-forming regions (SFRs) to the remains of supernovae (SNe). The detection of $\gamma$ rays at GeV and/or TeV energies from the molecular gas can serve to illuminate the high-energy particle production taking place in these objects located within the cloud and even in its vicinity. Indeed, at a first glance the spatial correspondence of a cloud with the $\gamma$-ray emission is susceptible to explain the latter via collisions of accelerated protons, for instance at supernova remnant (SNR) shock fronts, with target protons \citep[and maybe heavier nuclei,][]{banik+17} in the ambient matter. However, the recognition of this mechanism is not always straightforward since leptonic processes can also result in $\gamma$-ray emission via either bremsstrahlung or inverse Compton scattering produced by relativistic electrons.

Certainly, the number of $\gamma$-ray emitting sources detected at GeV energies by the \it Fermi \rm Gamma-ray Space Telescope is growing rapidly. More than 3000 sources were 
reported in the last catalogues presented by \citet{acero+15} and \citet{ackermann+16}. However, only for $\sim$35 of these sources a reliable counterpart originated in the radio emission from SNRs was found. Remarkably, approximately only half of these SNRs are well found in interaction with surrounding molecular gas.\footnote{See \url{http://www.physics.umanitoba.ca/snr/SNRcat/}.} 
As a counterpart to the GeV emission, several of the \it Fermi \rm sources were also detected at TeV energies\footnote{See \url{http://tevcat.uchicago.edu/}.}, and/or in the
radio/X-ray domains \citep[see for instance,][]{castro+13,acero+16}. However, regardless of the existence of a spatially coincident counterpart, the comprehension of the relative contribution of hadronic and leptonic processes responsible for $\gamma$-ray production remains unclear for a large fraction of cases \citep[e.g.][]{tanaka+11,pivato+13, abd18}.

Here we deal with the nature of the GeV $\gamma$-ray emission identified towards ($l$, $b$)$\simeq$(\dms{337}{\circ}{8},\dms{0}{\circ}{0}) with the Large Area Telescope (LAT) detector onboard the \emph{Fermi} satellite. The source known as 3FGL~J1838.6$-$4654 in the \emph{Fermi}-LAT source catalogue \citep[3FGL,][]{acero+15} lies in the GeV emitting region surveyed in this work.%
\footnote{The source 3FGL~J1838.6$-$4654 is catalogued as FL8Y~J1638.5$-$4654 in the preliminary version of the \emph{Fermi}-LAT 8-year source list, which will be replaced by the forthcoming 4FGL catalogue.} 
Given the spatial coincidence on the plane of sky, \citet{liu+15} considered natural a link between the GeV emission and the Galactic SNR~{\kes}. Moreover, using the properties of the molecular gas emission derived by \citet{zhang+15} in a restricted $\sim$7$^{\prime}$ $\times$ 4$^{\prime}$ area along the western rim of {\kes}, \citet{liu+15} proposed the hadronic interactions as the most feasible process to explain the production of the $\gamma$-ray emission in the GeV domain. 
The large-scale structure of the ambient matter in which {\kes} evolves was recently investigated in the companion paper presented by Supan et al. (2018a). 
As a result, on the basis of molecular and atomic line emission data, respectively from The Mopra Galactic Plane CO Survey \citep{burton+13} and the Southern Galactic Plane Survey \citep[SGPS,][]{mcclure+05}, along with mid-infrared \emph{Spitzer} \citep{churchwell09, carey09} information, the authors reported the discovery of the natal cloud ($\sim$26$^{\prime}$ in size, mass $M$$\sim$10-30~M$_{\odot}$) of the SNR and several star-forming regions around it. Now, in the light of the newly-determined physical conditions of the $\gamma$-ray emitting cloud, together with an updated set of \emph{Fermi}-LAT data and on radio observations, we re-analised the feasibility of hadronic and leptonic models to fit the broadband spectral energy distribution (SED) for the SNR/$\gamma$-ray source system.

\section{Observations and data analysis}\label{data-and-calculations}

\subsection{$\gamma$-ray observations}\label{data-fermi}
We analysed the field of {\kes} at $\gamma$-ray energies by using GeV data acquired with the {\Fermi} during $\sim$9 years of the mission since the beginning of it on August 08, 2008 to July 10, 2017 (that is, mission elapsed time between 239557417 and 521337605, respectively). This time interval greatly increases the observing time by $\sim$71\% with respect to the analysis previously presented in \citet{liu+15}, by adding $\sim$3.3~years of observations with the {\Fermi}. 
Our ojective is to obtain an up-to-date morphological and spectral picture of the GeV sky in the direction of the remnant {\kes} (i.e., towards the source 3FGL~J1838.6$-$4654), in order to investigate the nature of the observed $\gamma$-ray emission.

Events from a region of interest (ROI) with a radius of 10$^\circ$ centred in the source 3FGL~J1838.6$-$4654 ($l,b$ $\simeq$ \dms{337}{\circ}{798}, \dms{0}{\circ}{054}) were selected. The energy range 0.5-400~GeV was chosen for the spatial analysis, as a compromise between limited statistics and angular resolution. For the spatial analysis, the ROI was reconstructed with a pixel size of \dms{0}{\circ}{025}. 
Scientific products were obtained following a standard reduction chain, employing the {\it Science Tools} (ST) package version v10r0p5 and python, with the current version of event reconstruction (Pass~8, \citealt{atw13}) and the latest instrument response functions (IRFs P8R2\_SOURCE\_V6).%
\footnote{Details about specifications of the mission, instrumental issues, current event reconstruction framework, background models, etc, can be found at the {\Fermi} Science Support Center (FSSC) web page: \url{https://fermi.gsfc.nasa.gov/ssc/}.} 
In order to consider ``good'' photons for our analysis, we only kept events filtered according good-time-intervals (GTIs) and {\tt source} class events ({\tt evtype} = 3), which are indicated for point-source analysis. Additionally, we discarded photons coming from a zenith angle greater than 90$^\circ$ to minimize contamination due to $\gamma$ rays generated by interactions of cosmic rays with the upper atmosphere.

From the above mentioned filtering process we obtained a set of high-quality $\gamma$-ray data, which we analysed by means of the maximum likelihood technique \citep{mat96} implemented through the ST routine {\it gtlike}. We performed a binned likelihood analysis, dividing the 0.5-400~GeV range in 30 logarithmically-spaced energy bins. 
Background emission was modelled through the user-contributed script {\tt make3FGLxml.py} including point as well as extended sources in the ROI from the 3FGL catalogue along with the {\it gll\_iem\_v06} and {\it iso\_P8R2\_SOURCE\_V6\_v06} models for the diffuse Galactic and isotropic extragalactic components, respectively. 
The likelihood optimization procedure was carried out fixing spectral parameters (i.e., spectral index, normalization, etc) of the sources beyond 5$^{\circ}$ from the ROI centre, in order to ensure convergence. A first approximation of the spectral parameters was obtained by using the Minuit optimizer, which was then refined with the NewMinuit optimizer until convergence. Normalization of the Galactic and extragalactic backgrounds were left free.

\subsubsection{Spatial distribution}
\begin{figure}[!ht]
  \centering
\includegraphics[width=8cm]{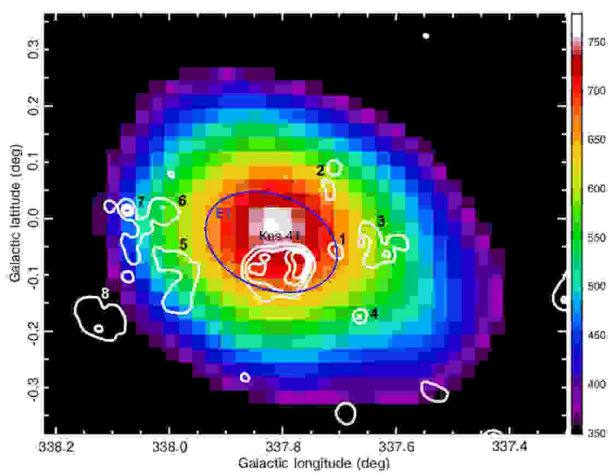}
\caption{New TS map in the region of SNR~{\kes} generated in the 0.5-400~GeV energy band according to the last update on this source analysed in this work. Contours delineate the radio emission at 843~MHz from SUMSS at levels 0.09, 0.3, 0.5, and 0.7~Jy~beam$^{-1}$. Labels correspond to {\kes} and the neighbouring HII regions (1:G337.711$-$0.056; 2:G337.711+0.089; 3:G337.622$-$0.067; 4:G337.667$-$0.167; 5:G337.978$-$0.144; 6:G338.011+0.022; 7:G337.686+0.137; 8:G338.114$-$0.193; \citealt{jones+12}). 
The ellipse E1 is the area employed in Supan et al. (2018a) to derive the properties of the interstellar medium, used to put constraints in the spectral fitting procedure presented in this work. Further details of the {\kes}'s environment are presented in Supan et al. (2018a).}
\label{gamma}
\end{figure}

In order to reveal the appearance of the GeV source in the field, we obtained a test-statistics (TS) map of the region by comparing the emission from the source with that of the constructed background model by using the tool {\it gttsmap}. 
The TS parameter is defined as $2\log(\mathcal{L}/\mathcal{L}_0)$, where $\mathcal{L}$ is the likelihood with source included and $\mathcal{L}_0$ corresponds to the null hypothesis (i.e., the sky subtracting the GeV excess of the source). It is a useful parameter to evaluate the statistical significance of a $\gamma$-ray excess, as it is related to the detection significance $\sigma$ of the source according to $\sigma \sim \sqrt{\mathrm{TS}}$. 
The new TS map for the region in which we are interested is presented in Fig.~\ref{gamma}, where the radio emission at 843~MHz obtained from the Sydney University Molonglo Sky Survey \citep[SUMSS,][]{bock+99} is depicted using white contours. For ease of reference, the HII regions in the field are labelled in Fig.~\ref{gamma} with numbers from 1 to 8. The TS angular distribution, with an overall significance of 27$\sigma$, is consistent with those presented by \citet{liu+15}.

\subsubsection{Spectral analysis}
We performed a new binned likelihood analysis of the data in order to study the spectral energy distribution of the source. In this case we restricted the range to 0.3-143~GeV, which was divided into 6 spectral bands, and we carried out a separated likelihood analysis for each of them. The extreme energies and the corresponding (logarithmic) central energy $E_c$, as well as the GeV fluxes for each interval are reported in Table~\ref{table_kes41_gamma_fluxes}, and represented in Figs.~\ref{leptonicM} and \ref{hadronic_model} (Sect.~\ref{modelling}). The distribution of the spectral points reveals a tendency for the flux of the source to fall below detection limits at an energy $\sim$50~GeV, suggesting a possible low-energy cutoff. This spectral shape may support the idea that events confidently associated with the source for energies above this value are scarce. For energies $\gtrsim$50~GeV only upper limits were obtained.

Assuming a power-law spectral shape, the resulting photon index is $\Gamma = 2.45 \pm 0.05$, which is consistent with the one obtained by \citet{liu+15}. The flux in the 0.3-143~GeV band is determined to be $(2.99 \pm 0.25) \times 10^{-8}$~ph~\cm{-2}~s$^{-1}$. The detailed modelling of the broadband spectral energy distribution from radio up to GeV energies is presented in Sect.~\ref{modelling}. 
The uncertainties reported in Table~\ref{table_kes41_gamma_fluxes} include statistical as well as systematic errors. Systematic errors are mainly associated with the propagation of inaccuracies in the effective area of the instrument, the point spread function, and with uncertainties related to the normalization of the Galactic diffuse background. 
Uncertainties associated with the effective area of the {\Fermi} vary across the energy range, reaching maximum values of $\sim$10\% at the highest energy considered in our study (\citealt{ack12}, see also the FSSC web site\footnote{\url{https://fermi.gsfc.nasa.gov/ssc/data/analysis/scitools/Aeff_Systematics.html}.}). 
On the other hand, systematics associated with the PSF are of the order of 5\% for energies below 100~GeV in our range, linearly increasing up to $\sim$20\% at higher energies.%
\footnote{\url{https://fermi.gsfc.nasa.gov/ssc/data/analysis/LAT_caveats.html}.} 
To estimate the systematics associated with the Galactic background, we followed \citet{abd09}, that is, after obtaining the best-fit values from the binned likelihood procedure, we varied the normalization of this background in $\pm$6\% in order to determine the departure of the spectral parameters under this artificially-fixed model from the previously determined best-fit values.

\begin{table}[h!]
\centering
\caption{Fluxes in the GeV range for the $\gamma$-ray source detected in correspondence with SNR {\kes}. For each flux, the first quoted error corresponds to the statistic uncertainty, while the second error depicts the sistematic one.}
\label{table_kes41_gamma_fluxes}
\begin{tabular}{cccc}\hline\hline
%-------------------------------------------------------------------------------------------
    $E_c$  &   Energy range  & Flux ($E^2 dN/dE$)                 & \multirow{2}{*}{TS} \\
    (GeV)  &       (GeV)     & ($10^{-12}$~erg~\cm{-2}~s$^{-1}$)  &                     \\\hline
%-------------------------------------------------------------------------------------------
     0.5   &  0.30 $-$ 0.84  & 13.8~$\pm$~2.3~$\pm$~4.4           & 133                 \\
     1.4   &  0.84 $-$ 2.34  & 11.5~$\pm$~1.1~$\pm$~2.8           & 163                 \\
     3.9   &  2.34 $-$ 6.55  &  8.4~$\pm$~0.8~$\pm$~2.1           & 150                 \\
    10.9   &  6.55 $-$ 18.3  &  2.9~$\pm$~0.7~$\pm$~1.3           & 28                  \\
    30.6   &  18.3 $-$ 51.2  &  $\leqslant0.9$ \tablefoottext{a}  & 8                   \\
    85.6   &  51.2 $-$ 143   &  $\leqslant1.6$ \tablefoottext{a}  & 3                   \\\hline
%-------------------------------------------------------------------------------------------
\end{tabular}
\tablefoot{
\tablefoottext{a}{Values corresponding to 95\% confidence level upper limits.}}
\end{table}

\section{The supernova remnant {\kes}}
\label{kes41-sec}
{\kes} ({\Gkes}) is a Galactic SNR classified, on the basis of the thermal X-ray emission detected towards its interior, as a member of the thermal-composite SNR class \citep{zhang+15}. Although it was proposed that the remnant could be the result of the collapse of a star no later than a B0-type, with a mass $\gtrsim$18~{\Msun} \citep{zhang+15}, the lack of a central compact object reported for this remnant brings into question the proposed nature of the progenitor star.

\begin{figure*}[ht!]
  \centering
\includegraphics[width=16cm]{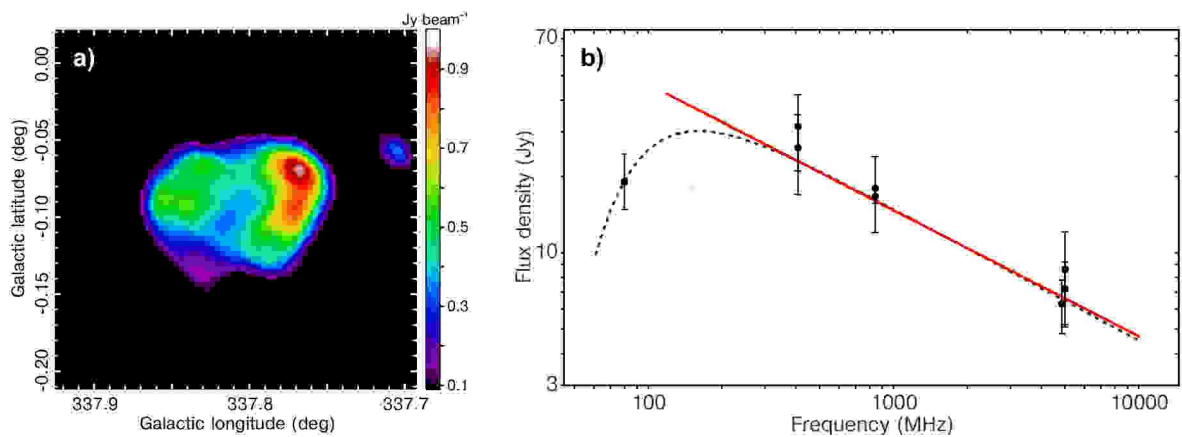}
  \caption{{\bfseries a)} SUMSS 843~MHz continuum emission from {\kes}, the beam size is $\sim$43{\as}~$\times$~59{\as} and the noise level is $\sim$0.015~{\Jyb}. A linear intensity scale was used for the representation. 
{\bfseries b)} Integrated radio spectrum of SNR {\kes} constructed with the flux densities measurements reported in Table~\ref{table_kes41_radio_fluxes}. Solid line corresponds to the weighted linear fit to the points excluding that at 80~MHz, which yields a global spectral index $\alpha$=$-0.54$$\pm$0.10 ($S_\nu \propto \nu^{\alpha}$). 
The same spectral index with an optical depth at 80~MHz $\tau_{\mathrm{80MHz}}$= 1.13$\pm$0.37 is obtained by fitting (dotted line) all of the measurements with a low frequency turnover in the spectrum around 100~MHz.}
  \label{kes41}
\end{figure*}

Here, we present a thorough look at {\kes} which as displayed in Fig.~\ref{gamma} is seen superimposed in projection on the GeV emission detected by {\Fermi}. A zoomed-in view of the remnant at 843~MHz from SUMSS is shown in Fig.~\ref{kes41}a. The remnant exhibits a slightly elongated radio shell of about $\sim$\dms{7}{\prime}{5}~$\times$~6{\am} in size (equivalent to an average size of 24~pc at the SNR distance of 12~kpc\footnote{The detection of the OH maser spot evidenced the encounter of the remnant with dense interstellar gas and served to place the remnant at a distance of $\sim$12~kpc \citep{koralesky+98}.}), brighter towards its Galactic western portion. 
In Supan et al. (2018a) it has been demonstrated that this bright emission is morphologically correlated with an enhancement in the molecular gas emission observed between $\sim$$-$63 and $\sim$$-$48~km~s$^{-1}$ (see Fig.~3 in that paper).

To calculate the global radio spectral index of the remnant we measured the integrated flux density of the source on images taken from public surveys at 843 and 5000~MHz, and constructed the spectrum shown in Fig.~\ref{kes41}b by adding these estimates to previously published flux density measurements for {\kes}. The list of the integrated flux density estimates is presented in Table~\ref{table_kes41_radio_fluxes}. The reported values, provided that the information about the primary calibrators is available, were brought onto the flux density scale of \citet{perley17}. 
From the spectrum it is evident that the flux density value at 80~MHz lies below the general trend of the data, which in principle may indicate the presence of thermal absorption along the line of sight. As only one point is a scarce evidence to define a spectral turnover at low frequencies, we first excluded the flux density value at 80~MHz and calculated the integrated spectral index in the radio band by a single weighted power-law least-squares fit (where $S_\nu \propto \nu^{\alpha}$) to the remaining data points in Fig.~\ref{kes41}b. The derived value of the integrated spectral index is $\alpha$= $-0.54$$\pm$0.10, which is fairly consistent with the estimation $\alpha$$\simeq$$-0.51$ presented by \citet{whiteoak-green-96} based only on integrated fluxes measured at 408 and 843~MHz. The integrated spectral index is also similar to that measured in other SNRs without a 
compact remnant in their interior. 
On the other hand, if the low frequency turnover is considered valid, a weighted least-squares fit to all the integrated flux densities shown in Fig.~\ref{kes41}b using a power law plus an exponential turnover at a fiducial frequency of 408~MHz ($S_{\nu}= S_{\mathrm{408}} \, (\nu / 408\,{\mathrm{MHz}})^{\alpha}\, \mathrm{exp}[{-\tau_{408} \, (\nu / 408 \, \mathrm{MHz}})^{-2.1}]$), produces a flux $S_{\mathrm{408}}$= 23.3$\pm$2.6~Jy and an optical depth $\tau_{\mathrm{408}}$=0.037$\pm$0.012 indicative of a non significant absorption level at this frequency. The free-free optical depth at 80~MHz is $\tau_{\mathrm{80}}$=1.13$\pm$0.37 ($\tau_{\mathrm{80}}=\tau_{\mathrm{408}}\, [80/408]^{-2.1}$). The global spectral index has the same value derived when absorption from thermal ionised gas along the line is not present. Certainly, more flux density measurements at frequencies below 100~MHz are needed to clarify the presence of the spectral turnover.

\begin{table}[h!]
\centering
\caption{Flux densities used to construct the global radio continuum spectrum of {\kes}.}
\label{table_kes41_radio_fluxes}
\begin{tabular}{r D{?}{\,\pm\,}{3.3} l}\hline\hline
%-----------------------------------------------------------------------------------
 Frequency &   \multicolumn{1}{c}{Flux density}  & \multirow{2}{*}{References}    \\
   (MHz)   &       \multicolumn{1}{c}{(Jy)}      &                                \\\hline
%-----------------------------------------------------------------------------------
     80    &                19.9 ? 5.0           &  \citet{dulk-slee-72}          \\
    408    &                26.6 ? 9.2           &  \citet{shaver-goss-70}        \\
    408    &                33.1 ? 11.0          &  \citet{dulk-slee-72}          \\
    843    &              16.7 ? 1.0             &  This work\tablefoottext{a}    \\
    843    &                18.0 ? 6.0           &  \citet{whiteoak-green-96}     \\
   4850    &               6.1 ? 1.4             &  \citet{PMN-2-94}              \\
   5000    &               7.3 ? 2.0             &  \citet{dulk-slee-72}          \\
   5009    &               8.6 ? 3.5             &  This work\tablefoottext{a}    \hfil\\\hline
%-----------------------------------------------------------------------------------
\end{tabular}
\tablefoot{
\tablefoottext{a}{Flux densities from the 843~MHz SUMSS image \citep{bock+99} and 5009~MHz Parkes telescope survey \citep{haynes+78}.}}
\end{table}

\section{Physical scenarios for the GeV $\gamma$ rays}
\label{modelling}
It is believed that the $\gamma$-ray radiation produced in supernova remnants originates by the interaction of accelerated hadrons and/or leptons with surrounding ambient matter and radiation. This high energy $\gamma$-ray flux is enhanced in presence of dense MCs \citep[see][for a review]{slane+15}.

In \citet{liu+15} the $\gamma$-ray flux of {\kes} is modelled considering leptonic and hadronic scenarios. Assuming that the $\gamma$ rays originate by inverse Compton scattering 
of accelerated electrons with the low energy photons of the Cosmic Microwave Background, they found that the required energy in accelerated electrons to reproduce the observed flux is of the order of $1.3\times10^{51}$ erg. This result is about one order of magnitude larger than the canonical value of $10^{50}$ erg, which corresponds to a fraction of 10\% of the typical energy released in a supernova explosion. They also developed hadronic models compatible with the observed flux and also with the canonical values of the energy that goes to the accelerated particles in a supernova explosion.

In the current work, we take the properties derived for the large cloud into account (Supan et al. 2018a) and use, for the first time, a broadband spectrum including observations at radio energies to review the plausibility of a hadronic and leptonic origin of the $\gamma$-ray flux in the region of the SNR~{\kes}.

The energy distribution of the accelerated particles in both, the hadronic and leptonic, scenarios considered here is assumed to be a power law with an exponential cutoff,
\begin{equation}\label{Spec}
  \frac{dN_{a}}{dE} = K_{a}\ E^{-\Gamma_{a}} \exp\left(-E/E_{cut_{a}} \right),
\end{equation}
where $a=\{e,p\}$ indicates the particle species (electrons or protons), $K_{a}$ is a normalization constant, $\Gamma_{a}$ is the spectral index, and $E_{cut_{a}}$ is the cutoff energy. 

Let us first consider a model in which leptons are the component responsible for the $\gamma$-ray flux. In this type of model the radio flux is due to synchrotron radiation emitted by the propagation of the accelerated electrons in the ambient magnetic field. 
On the other side, the $\gamma$-ray radiation is generated mainly by the interaction of the accelerated electrons with the ambient matter, via the nonthermal bremsstrahlung process, and with the ambient radiation field, via inverse Compton. We refer to \citet{aharonian+10} and references therein for calculation details of the synchrotron emissivity. Regarding the inverse Compton and the nonthermal bremsstrahlung emission, we followed the method given in \citet{jones-68} for the former, while the method presented in \citet{baring+99} was used to model the electron-electron bremsstrahlung interaction and in \citet{koch-motz-59} and \citet{sturner+97} for the electron-ion bremsstrahlung process.

In modelling the spectrum from radio to $\gamma$-ray energies, we consider the radio observations reported in Table~\ref{table_kes41_radio_fluxes} and the updated {\Fermi} data presented in Sect.~\ref{data-fermi}. The chi-square used to fit the data is given by,
\begin{equation}\label{chi2}
  \chi^2(\vec{\theta})=\sum_{i=1}^N \frac{(J_i-J(E_i;\vec{\theta}))^2}{\sigma_i^2} + \frac{(n_p-\bar{n}_p)^2}{\sigma[n_p]^2},
\end{equation} 
where $J_i$ corresponds to the observed flux at energy $E_i$ with an uncertainty $\sigma_i$, and $J(E_i;\vec{\theta})$ is the model under consideration with parameters $\vec{\theta}=(\Gamma_e,K_e,E_{cut_e},B,n_p )$. Here, $B$ is the ambient magnetic field intensity and $n_p$ the proton density. We notice that the last term in $\chi^2(\vec{\theta})$ includes the uncertainty on the determination of the proton density. 
The values of $\bar{n}_p$ and $\sigma[n_p]$ correspond to the E1 region enclosing the SNR~{\kes}, which is drawn in Fig.~\ref{gamma}. A comprehensive analysis of the physical conditions in this region is presented in Supan et al. (2018a). The inclusion of the mean proton density as one of the fitting parameters makes the fit to take into account its correlation with the other fitting parameters. 
Moreover, the covariance matrix of the fit depends on the estimated mean proton density. Therefore, the mean proton density uncertainty is properly propagated (including correlations) in subsequent calculations that are based on the fitting parameters like, for instance, the total energy in accelerated electrons corresponding to the model.

\begin{figure}[!ht]
  \centering
 \includegraphics[scale=0.175]{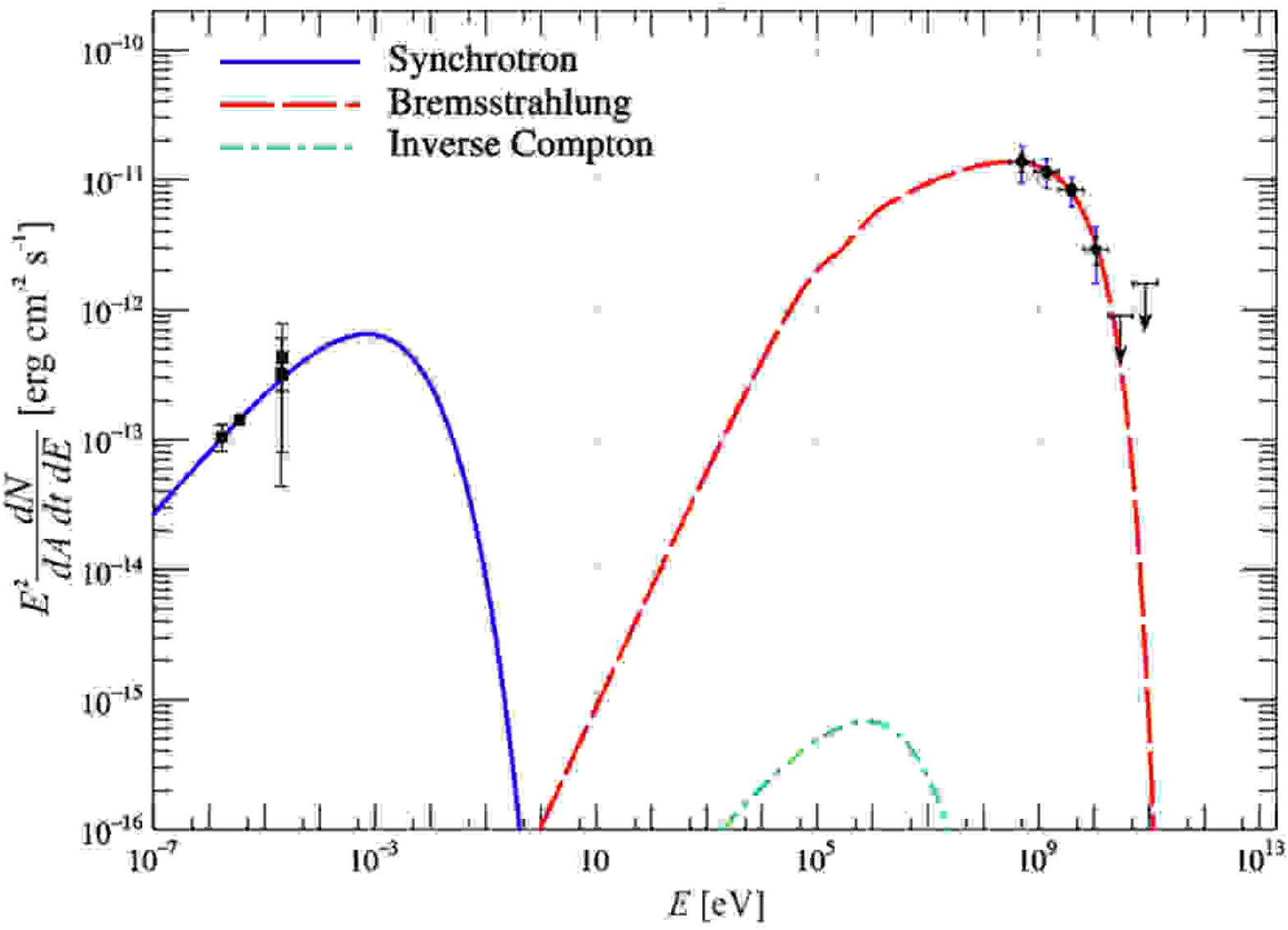}
  \caption{Spectral energy distribution of {\kes}. The curves correspond to the leptonic models fitted to the experimental data. At GeV energies the statistical uncertainties are indicated by black bars while the systematic ones are represented in blue. The spectral index of the accelerated electrons used for the fit is $\Gamma_e=2$. Upper limits for the GeV fluxes correspond to a 95\% confidence level, obtained for TS values $<$ 9.}
  \label{leptonicM}
\end{figure}

The spectral index of the accelerated electrons energy distribution is fixed during the $\chi^2$ minimization procedure. Figure~\ref{leptonicM} shows the fit of the experimental data for $\Gamma_e=2$, this value is motivated by the prediction of the first order Fermi mechanism and it is contained in the one-sigma region of the radio data fit performed in Sect.~\ref{kes41-sec}. 
As Fig.~\ref{leptonicM} shows, the nonthermal bremsstrahlung process is the dominant contribution to the $\gamma$-ray part of the flux. The inverse Compton component is several orders of magnitude smaller than the corresponding one to nonthermal bremsstrahlung. The fitted parameters are $\log(\hat{E}_{cut_e}/\textrm{eV}) = (10.09 \pm 0.08)$, $\hat{B} = (103 \pm 24)\ \mu\textrm{G}$, and $\hat{n}_p = ( 957 \pm 329 )$ cm$^{-3}$ (the hat is included to emphasize the maximum likelihood estimates character of the parameters). 
The obtained cutoff energy of the accelerated electrons is quite small. This is due to the suppression present in the $\gamma$-ray data. It is worth mentioning that, the high value of the proton density requires a smaller number of accelerated electrons to reproduce the $\gamma$-ray data and then, a high value of the magnetic field is required to reproduce the radio data.

The energy in accelerated electrons is calculated by using Eq.~(\ref{Spec}). For the spectral index $\Gamma_e=2$ the following value is obtained, 
\begin{equation}
  \mathcal{E}_\mathrm{e} \left(\Gamma_e=2\right)= \left(5.0 \pm 1.8 \right)\times 10^{48}\ \textrm{erg}. 
\end{equation} 
The total energy in accelerated electrons is smaller than the canonical value of $10^{50}$ erg. The maximum energy in accelerated electrons is obtained by fixing the spectral index in $\Gamma_e=2.28$, which is the upper limit of the one-sigma region obtained from the fit of the radio data (see Sect.~\ref{kes41-sec}). In this case, a good fit is still obtained with the following value for the total energy,
\begin{equation}
  \mathcal{E}_\mathrm{e} \left(\Gamma_e=2.28\right)= \left(7.1 \pm 2.5 \right)\times 10^{48}\ \textrm{erg}. 
\end{equation} 
This value is still smaller than $10^{50}$ erg. Therefore, a leptonic origin of the $\gamma$-ray flux is compatible with the present data. 

Let us now consider the hadronic model. In this type of models the $\gamma$ rays originate in the decay of particles, mainly neutral pions, generated in the interactions of accelerated protons with ambient protons. As mentioned before, the energy spectrum of the accelerated protons is assumed to be a power law with an exponential cutoff (see Eq.~(\ref{Spec})). The $\gamma$-ray flux at Earth can be written as
\begin{equation}
\label{eqn_gflux}
  J(E_\gamma) = \frac{c}{4\pi\ d^2}\ n_\mathrm{p} \int_{E_\gamma}^\infty dE_p \; \frac{dN_p}{dE_p}(E_p) %
                \; \frac{d\sigma}{dE_\gamma}(E_\gamma,E_p),
\end{equation}
where $d$ is the distance to the source, $d\sigma/dE_\gamma$ is the differential cross-section in proton-proton collisions resulting in the $\gamma$-ray emission, and $c$ is the speed of light. 

The differential cross section of $\gamma$ rays originated in proton-proton interactions has been extensively studied in the literature. A comprehensive study have been performed in \citet{Kafexhiu-14}, in which a parametrization of the $\gamma$-ray energy spectrum, originated in proton-proton collisions, has been obtained. The projectile proton energy range of the parametrization starts at the kinematic threshold reaching a maximum value of 1~PeV. The differential cross section at low proton energies ($\lesssim 3$~GeV) is obtained from a compilation of experimental data, while at high proton energies it is derived from Monte Carlo simulations. It is very well known that the hadronic interactions at the 
highest energies are unknown. However, there are models that extrapolate low energy accelerator data to the highest energies. 
The parametrization in \citet{Kafexhiu-14} includes several options for the high energy part, in fact it has been performed for the hadronic interaction models implemented in Geant 4.10.0 \citep{geant4}, PYTHIA 8.18 \citep{pythia}, Sibyll 2.1 \citep{sibyll}, and QGSJET01 \citep{qg01}. In this work, we used this parametrization of the proton-proton collisions 
resulting in the $\gamma$-ray emission with the PYTHIA 8.18 option for the high-energy part. 
It is worth mentioning that one of the most important aspects of the new parametrization is the detailed description of the $\gamma$-ray spectrum at low proton energies, which represents an important improvement over earlier approaches.

\begin{figure}[!ht]
  \centering
  \includegraphics[scale=0.175]{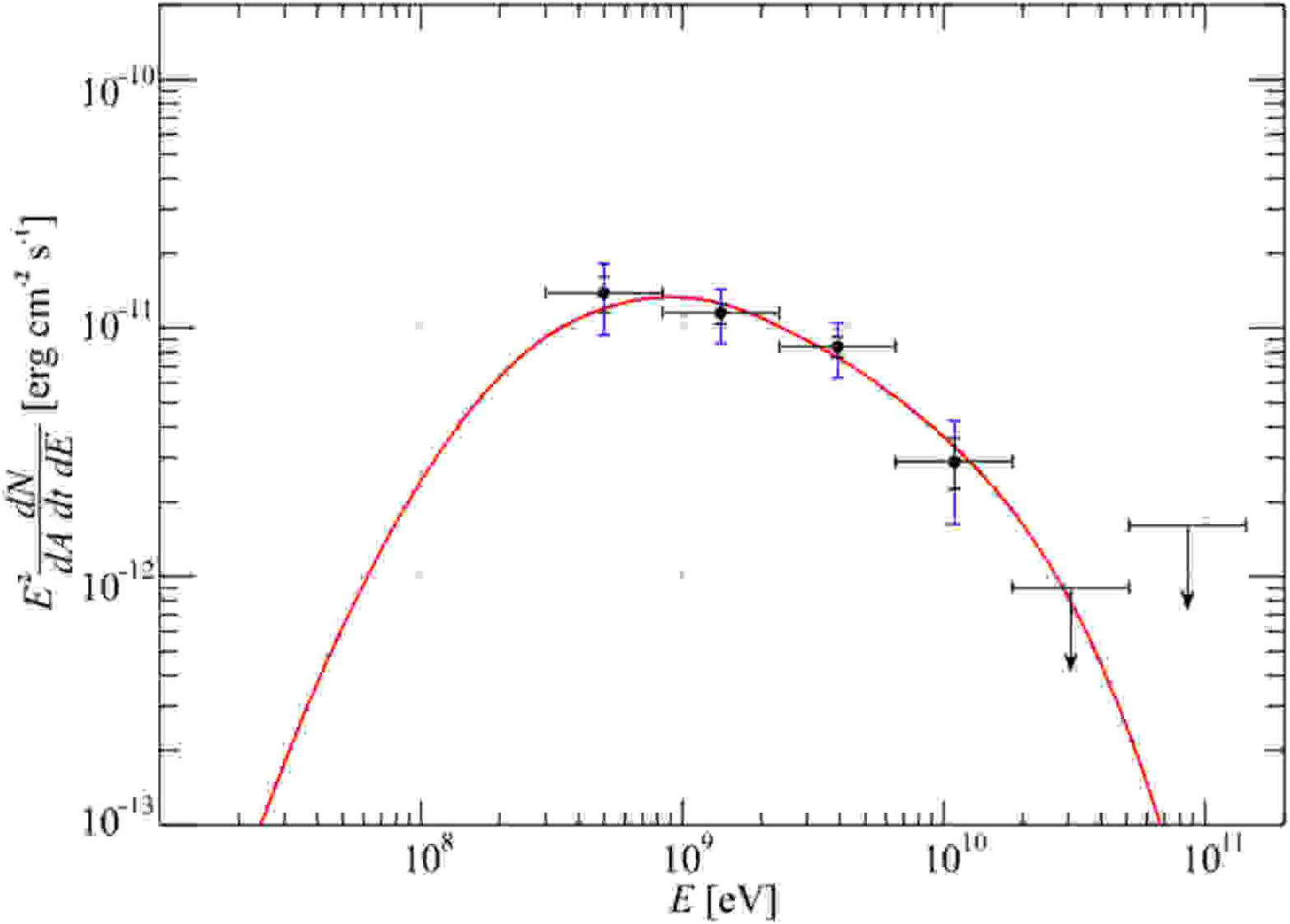}
  \caption{
{\Fermi} $\gamma$-ray spectrum in the region of SNR~{\kes}. The curve corresponds to the hadronic model fitted to the updated experimental data with a fixed spectral index $\Gamma_p=2$. The statistical and systematic errors are represented by black and blue bars, respectively. Upper limits for the flux correspond to a 95\% confidence level, obtained for TS values $<$ 9.}
  \label{hadronic_model}
\end{figure}

Figure~\ref{hadronic_model} shows the fit of the $\gamma$-ray part of the SNR~{\kes} energy spectrum observed at Earth. In this case we considered the value of the proton density, $n_{\mathrm{p}}$$=950\pm330$~cm$^{-3}$, in the SNR surroundings, derived inside the E1 region (see also analysis in Supan et al. 2018a). 
In addition, for this part of our analysis the spectral index $\Gamma_p$ is fixed during the fitting procedure. As in the case of the leptonic model we consider the canonical value $\Gamma_p=2$ and the values in the one-sigma region corresponding to the fit of the radio data. The cutoff energy obtained for $\Gamma_p=2$ is $\log(\hat{E}_{cut_p}/\textrm{eV}) = ( 10.70 \pm 0.09 )$.

The energy in accelerated protons calculated for the ambient proton density measured in the E1 region and for $\Gamma_p=2$ is given by, 
\begin{equation}
  \mathcal{E}_\mathrm{p}\left(\Gamma_p=2\right) = \left(1.32 \pm 0.47 \right)\times 10^{49}\ \textrm{erg}, 
\end{equation} 
which is smaller than the canonical value of $10^{50}$ erg. Moreover, considering the spectral index $\Gamma_p=2.28$,
the upper limit of the one sigma region for the fit of the radio data, the maximum value obtained for the energy in accelerated
protons is
\begin{equation}
  \mathcal{E}_\mathrm{p} \left(\Gamma_p=2.28\right)= \left(1.57 \pm 0.56 \right)\times 10^{49}\ \textrm{erg}, 
\end{equation} 
which, also in this case, is smaller than the canonical value.

It is noteworthy that from the leptonic and hadronic models developed in this section in which $\Gamma_e=\Gamma_p=2$, a ratio $\tilde{K}_{ep}(E)$$\sim 0.06$ between the differential number of accelerated electrons and the differential number of accelerated protons is obtained for $E \ll 10^{10}$~eV. This ratio, defined as \citep{Merten:17}
\begin{equation}
  \tilde{K}_{ep}(E)=\frac{\mathop{\displaystyle \frac{dN_e}{dE}(E)}}{\mathop{\displaystyle \frac{dN_p}{dE}(E)}},
\end{equation}
is crucial in modelling the nonthermal emission produced in cosmic ray sources. In our leptonic model it is assumed that the $\gamma$-ray part of the flux coming from the hadronic component is much smaller than the one corresponding to bremsstrahlung. Therefore, a value of $\tilde{K}_{ep} \sim 0.6$ is required in order to obtain a contribution from the proton component one order of magnitude smaller than the one corresponding to the bremsstrahlung process. This value is more than one order of magnitude larger than the one inferred from cosmic rays observations \citep{Merten:17}, which is generally assumed to be of the order of $0.01$. However, we recall that such a number represents an average value and it is not possible to know the differential particle number ratio in each individual source that contributes to the observed average. The canonical value of $\tilde{K}_{ep}$ can be obtained theoretically by assuming that the number of protons and electrons that are accelerated are the same and that the spectral indexes of both components after acceleration are also the same, taking a value of $\sim 2.2$. 
As discussed in \cite{Merten:17}, $\tilde{K}_{ep}$ is affected by several factors which can strongly modify its canonical value. In particular, there are evidences, both from a theoretical and observational point of view, which show that the spectral indexes of both components can be different. In that work, it is also shown that even for smaller values of the difference between the two spectral indexes a large variation of $\tilde{K}_{ep}$ is obtained. For instance, for $\Delta \Gamma=\pm 0.3$ the differential particle ratio varies in such a way that $10^{-5} < \tilde{K}_{ep} < 10$. Therefore, models which requires large values of $\tilde{K}_{ep}$ even of order one cannot be discarded (see for instance \citealt{HESS:16}). 
We cautiously note that the justification of a negligible contribution of the bremsstrahlung process in the leptonic model of \cite{liu+15} is based on the assumption of the canonical value of $\tilde{K}_{ep}$.

The analysis using the {\Fermi} data presented in this paper shows that the $\gamma$-ray part of the spectrum of {\kes} can be dominated by either accelerated leptons producing bremsstrahlung emission or accelerated hadrons that generate $\gamma$-ray emission by interacting with the surrounding ambient matter. The $\gamma$-ray flux can be properly fitted by these two types of models and also the total energy in accelerated particles is, in all cases, smaller than the canonical 10\% of the typical energy released in a supernova explosion.

\section{Concluding remarks}
This paper is the second in a series focused on the multiwavelength analysis of the counterparts to the unidentified $\gamma$-ray emission detected at GeV energies towards the SNR~{\kes}. 
Motivated by the new results recently presented in Supan et al. (2018a) revealing the natal molecular cloud of {\kes}, we performed, for the first time, a global assessment of all available data from radio to $\gamma$ rays in order to determine the relative contribution of the hadronic and leptonic processes to the overall spectrum. To do this, we re-analised the \emph{Fermi}-LAT data to study the $\gamma$-ray radiation spatially coincident (in projection) with the recently unveiled large-scale molecular material in the region of the SNR~{\kes}. 
Using the leptonic models analysed in this work, we demonstrated that if the main contribution to the $\gamma$-ray emission comes from the SNR then, both radio and $\gamma$-ray data can be successfully modelled by synchrotron radiation and bremsstrahlung mechanism, respectively. 
A leptonic contribution from the inverse Compton emission to explain the production of the emission at GeV energies can be easily excluded. We also modelled the {\Fermi} spectral data by a hadronic scenario considering the molecular, atomic, and ionised interstellar gases and found that \rm hadronic interactions in a region relatively close to {\kes} provide 
a viable mechanism for explaining the observed $\gamma$-ray emission, which qualitatively agree with the results of \citet{liu+15}.

\bibliographystyle{aa}
 \bibliography{mopra}

\begin{acknowledgements}
The authors wish to thank the anonymous reviewer since this manuscript was significantly improved by her/his comments. 
G. Castelletti and A.~D. Supanitsky are members of the {\it Ca\-rre\-ra del Investigador Cient\'{\i}fico} of CONICET, Argentina. L. Supan is a post-doc Fellow of CONICET, Argentina. 
This research was partially supported by grants awarded by ANPCYT (PICT~1759/15) and the University of Buenos Aires (UBACyT 20020150100098BA), Argentina.
\end{acknowledgements}

\end{document}